\documentclass[10pt,superscriptaddress,aps,prapplied,twocolumn,nofootinbib]{revtex4-2}
\usepackage{amsmath,amssymb,amsthm}
\usepackage{multirow}
\usepackage{tabularx}
\usepackage[breaklinks=true,colorlinks,citecolor=blue,linkcolor=blue,urlcolor=blue]{hyperref}
\usepackage[pdftex]{graphicx}
\usepackage{times,txfonts}
\usepackage{braket}
\usepackage{color}
\usepackage{natbib}
\usepackage{url}
\usepackage{xurl}
\usepackage{makecell}
\usepackage{bm}
\usepackage{amsmath,blkarray}
\usepackage{mathtools}
\usepackage{makecell}
\usepackage[normalem]{ulem}
\usepackage{latexsym}
\usepackage{tabularx, booktabs}
\usepackage{graphics,epstopdf}
\usepackage{graphicx}
\usepackage[linesnumbered,lined,ruled,vlined]{algorithm2e}
\usepackage{amsfonts}
\usepackage{tikz}
\usepackage{graphicx}
\usepackage{subfigure}
\usepackage{siunitx}

\usepackage{color}
\DeclareMathAlphabet{\mymathbb}{U}{BOONDOX-ds}{m}{n}

\newcommand{\be}{\begin{equation}}
\newcommand{\ee}{\end{equation}}
\newcommand{\ba}{\begin{eqnarray}}
\newcommand{\ea}{\end{eqnarray}}

\usepackage{multirow}
\usepackage{appendix}
\usepackage{verbatim}
\usepackage{orcidlink}

\begin{document}

\title{Digitized Counter-Diabatic Quantum Optimization for Bin Packing Problem}

\author{Ruoqian Xu$^{\orcidlink{0009-0005-3242-9674}}$}
\affiliation{Instituto de Ciencia de Materiales de Madrid (CSIC), Cantoblanco, E-28049 Madrid, Spain}

\author{Sebastián V. Romero$^{\orcidlink{0000-0002-4675-4452}}$}
\affiliation{Department of Physical Chemistry, University of the Basque Country UPV/EHU, Apartado 644, 48080 Bilbao, Spain}
\affiliation{Kipu Quantum GmbH, Greifswalderstrasse 212, 10405 Berlin, Germany}

\author{Jialiang Tang$^{\orcidlink{0009-0001-6420-6492}}$}
\affiliation{Instituto de Ciencia de Materiales de Madrid (CSIC), Cantoblanco, E-28049 Madrid, Spain}

\author{Yue Ban$^{\orcidlink{0000-0003-1764-4470}}$}
\email{yue.ban@csic.es}
\affiliation{Instituto de Ciencia de Materiales de Madrid (CSIC), Cantoblanco, E-28049 Madrid, Spain}

\author{Xi Chen$^{\orcidlink{0000-0003-4221-4288}}$}
\email{xi.chen@csic.es}
\affiliation{Instituto de Ciencia de Materiales de Madrid (CSIC), Cantoblanco, E-28049 Madrid, Spain}

\date{\today}

\begin{abstract}
The bin packing problem, a classical NP-hard combinatorial optimization challenge, has emerged as a promising candidate for quantum computing applications. In this work, we address the one-dimensional bin packing problem (1dBPP) using a digitized counter-diabatic quantum algorithm (DC-QAOA) that incorporates counter-diabatic (CD) driving to reduce quantum resource requirements while maintaining high solution quality, outperforming traditional methods such as QAOA. 
We evaluate three ansatz schemes—DC-QAOA, CD-inspired ansatz, and CD-mixer ansatz—each integrating CD terms with distinct combinations of cost and mixer Hamiltonians, resulting in different DC-QAOA variants.  Among these, the CD-mixer ansatz demonstrates superior performance, showing robustness across various iteration counts, layer depths, and Hamiltonian steps, while consistently producing the most accurate approximations to exact solutions.
To validate our approach, we solve a 10-item 1dBPP instance on the IBM quantum computer, optimizing circuit structures through simulations.
Despite constraints on circuit depth, the CD-mixer ansatz achieves high accuracy and likelihood of success. These findings establish DC-QAOA, particularly the CD-mixer variant, as a powerful framework for solving combinatorial optimization problems on near-term quantum devices.
\end{abstract}

\maketitle

\section{Introduction}
\label{Sec1:intro}

In computational science, optimization problems pose significant challenges across various applications, including logistics management, supply chain optimization, and resource allocation in computer systems. The bin packing problem (BPP)~\cite{Garey1981}, a well-known NP-hard problem~\cite{10.5555/574848}, captures this complexity with its straightforward formulation and intricate solution process. Broadly, the BPP involves allocating items into a minimal number of bins while satisfying specific constraints and objectives. Traditional approaches for solving BPP have included brute force methods~\cite{LEVIN2022646}, classical heuristics~\cite{vazirani2013approximation, johnson1973near}, machine learning algorithms~\cite{s23156928, 8923909}, and evolutionary algorithms~\cite{RePEc:ids:ijmore:v:6:y:2014:i:6:p:732-751, GAN,Zhang2011}. More recently, quantum computing approaches such as quantum annealing~\cite{deandoin2022comparative, de2022hybrid, 1188899} and other quantum algorithms~\cite{romero2023solving, v2023hybrid} have emerged as innovative solutions, leveraging quantum mechanics to tackle optimization problems. However, these quantum approaches still face limitations in scalability and efficiency, particularly for large-scale and high-dimensional instances of the BPP. The complexity of the BPP increases further with the dimensionality of bins and items, making it a compelling benchmark for exploring new computational paradigms and optimization techniques.

The simplest yet most versatile variant of BPP is given by the one-dimensional BPP (1dBPP)~\cite{Munien2021metaheuristic}, which has been mostly used for balanced air cargo loading~\cite{1188899}, logistics~\cite{10.1007/978-3-030-59747-4_22}, and task scheduling~\cite{WITTEMAN2021365}, among others. In the case of the two-dimensional BPP (2dBPP)~\cite{LODI2002379}, apart from logistics~\cite{G2001193} and transportation-related use cases~\cite{LEUNG2011205}, it also has practical applications for cutting processes in fabrics~\cite{9093914,PARRENO2020378}. Meanwhile, the three-dimensional BPP (3dBPP)~\cite{martello2000three} is the most studied one because of its closeness to more realistic cargo loading scenarios~\cite{v2023hybrid}. Among these variants, 1dBPP stands out as a suitable testbed for quantum optimization algorithms on current quantum computers. Its unique combination of simplicity and relevance in transportation and logistics
makes it both a challenging optimization problem and an excellent benchmark for evaluating algorithm performance on quantum hardware. The simplified 1dBPP unlocks the potential of quantum computing to address industrial-scale challenges, paving the way for integrating current and emerging quantum platforms into real-world applications.

Previous studies have investigated the integration of quantum protocols to address the 1dBPP, particularly using quantum annealing. In Ref.~\cite{de2022hybrid}, a simplified 1dBPP model was introduced, followed by benchmarking efforts~\cite{deandoin2022comparative} that highlighted the comparative advantages of the approach. Further experimental progress~\cite{Cellini2024} has demonstrated that current quantum devices are capable of finding optimal 1dBPP solutions. Similar quantum methods have been extended to other BPP variants, such as the 3dBPP~\cite{v2023hybrid}, job scheduling~\cite{Carugno2022}, and the traveling salesman problem
\cite{dalal2024digitizedcounterdiabaticquantumalgorithms}. The emergence of variational quantum algorithms (VQAs)~\cite{cerezo2021variational}, including the quantum approximate optimization algorithm (QAOA)~\cite{doi:10.1126/science.1057726},   has shown significant promise, with experimental validation~\cite{Guerreschi2019}. To further enhance QAOA’s efficiency, several advancements have been developed, focusing on faster convergence and reduced circuit depth. Techniques such as adaptive bias QAOA and ADAPT-QAOA optimize the mixer Hamiltonian to minimize circuit complexity~\cite{PhysRevResearch.4.023249, zhu2022adaptive}, while warm-start QAOA leverages problem-specific initial parameters to boost performance~\cite{egger2021warm}. Among these, the digitized counter-diabatic QAOA (DC-QAOA) inspired by shortcuts to adiabaticity (STA) \cite{PhysRevApplied.15.024038} has demonstrated remarkable success in addressing combinatorial problems such as MaxCut~\cite{wurtz2022counterdiabaticity,chandarana2022meta,PhysRevResearch.4.013141}, portfolio optimization~\cite{PhysRevResearch.4.043204}, protein folding \cite{PhysRevApplied.20.014024} and molecular docking~\cite{PhysRevApplied.21.034036}.

In this article, we present the first in-depth exploration of the 1dBPP using QAOA enhanced by counter-diabatic (CD) protocols. While quantum computing has shown considerable potential in various theoretical and experimental contexts, its practical application to specific problems like BPP remains underexplored. We compare the performance of DC-QAOA and QAOA, focusing on three DC-QAOA variants: the original DC-QAOA, CD-inspired ansatz, and CD-mixer ansatz. These variants integrate CD terms with distinct combinations of cost and mixer Hamiltonians, and we analyze their robustness across various iteration counts, layer depths, and Hamiltonian steps. This work aims to address this gap by offering a comprehensive investigation of parameterized circuits for solving the 1dBPP. Through this, we seek to assess the performance of the quantum algorithm in comparison to to traditional methods, while also exploring its scalability and practical applicability in real-world scenarios.

The remainder of this work is organized as follows. We begin with a concise introduction to the traditional and simplified model of 1dBPP in Sec. \ref{sec: bpp}, laying the foundation for a deeper exploration of the methodology. In Sec. \ref{sec: algorithm}, we present the quantum framework and algorithms employed in our study. Through the numerical simulations and practical implementation on IBM quantum computer, Sec. \ref{sec: results} and Sec. \ref{sec: exp} provide our findings, including a comprehensive analysis with classical approaches, highlighting the advantages and limitations of quantum methods. Finally, in Sec. \ref{sec: conclusion}, we conclude the broader implications of our results and discuss the potential future impact of quantum computing on solving optimization challenges.

\section{Preliminaries}
\label{sec: bpp}

\subsection{Classical formulation}

The 1dBPP is often considered a prototype optimization problem, involving the task of packing $ n $ items with varying weights into $m$ bins under certain constraints. The primary objective is to minimize the number of bins used while ensuring all items are packed. Each bin has a fixed capacity $\mathcal{C}$, and each item $i$ has a weight $\omega_{i}$, which must not exceed the bin’s capacity. Thus, the 1dBPP can be formulated as a binary-constrained optimization problem as follows:
\begin{subequations}
\begin{align}
  \min\quad & \sum_{j=1}^{m}y_{j}, \label{eq:bpp-1}\\
  \text{s.t.}\quad & \sum_{i=1}^{n} \omega_{i} x_{i,j} \leq \mathcal{C}, \quad \forall j \in [1,m], \label{eq:bpp-2}\\
   & \sum_{j=1}^{m} x_{i,j} = 1, \quad \forall i \in [1,n],\label{eq:bpp-3}
\end{align}
\end{subequations}
where $ y_j $ represents the binary variable indicating whether bin $ j $ is used, and $ x_{i,j} $ denotes the binary variable that specifies if item $ i $ is placed in bin $ j $. Eqs.~\eqref{eq:bpp-2} and \eqref{eq:bpp-3} ensure that the total weight of items in any bin does not exceed the maximum capacity $ \mathcal{C} $ and that each item is packed exactly once. Given these constraints, Eq.~\eqref{eq:bpp-1} minimizes the total number of active bins in use.

The most straightforward method for solving the 1dBPP is through brute force methods, 
which systematically evaluates all possible combinations to determine the exact solution set
$\mathcal{F}$. However, this approach has an exponential time complexity of $m^n$, making it impractical for larger problem sizes due to the excessive computational resources required. In contrast, classical heuristic algorithms, such as those with linear or polynomial time complexities (e.g., $n$ or $n \log{m}$), provide more efficient 
solutions~\cite{CoffmanJr.2013,9300141,johnson1973near} at the cost of reduced accuracy compared to $\mathcal{F}$~\cite{LEVIN2022646}.
Recent advancements in machine learning have introduced innovative optimization strategies for tackling the 1dBPP. Techniques such as transfer learning~\cite{montanez2024transfer}, genetic algorithms~\cite{GAN}, and reinforcement learning~\cite{s23156928}, have been employed to leverage data-driven insights for more efficient packing solutions.

\subsection{Hamiltonian and quantum encoding}

Quantum computing offers innovative solutions to optimization problems, such as the 1dBPP, by reformulating classical constraints into a qubit-compatible representation. However, transforming the constraint~\eqref{eq:bpp-2} into a Hamiltonian introduces numerous variables, a challenge that we aim to mitigate~\cite{de2022hybrid}. Drawing inspiration from prior studies~\cite{montanezbarrera2023unbalanced, deandoin2022comparative}, we propose a function that satisfies the constraint in Eq.~\eqref{eq:bpp-2} for a single bin. The objective function can be expressed as follows:
\begin{eqnarray}
\label{eq:idea}
f\left(\sum_{i=1}^{n}\omega_{i}x_{i}\right)=
    \left\{
     \begin{array}{ll}
    \infty & \text{if }\sum_{i=1}^{n}\omega_{i}x_{i} > \mathcal{C}, \\
    0 & \text{otherwise}.
      \end{array}
  \right.
\end{eqnarray}
Here $ x_i $ is a binary variable that equals 1 if item $ i $ is placed in the bin, and 0 otherwise. The objective is to compute $ \sum_{i=1}^{n} \omega_i x_i $, which represents all feasible packing solutions within a single bin.

Due to current quantum device limitations, a stepwise Hamiltonian implementation is impractical. Instead, we construct a set of quadratic Hamiltonians whose ground states correspond to different weights. The Hamiltonian is given by: 
\begin{equation}
\label{eq:h_p_2_mod}
    H_{\text{binary}} = \mathcal{A}\left( \sum_{i=1}^{n}\omega_{i}x_{i} -\mathcal{C}\right) + \mathcal{B}\left( \sum_{i=1}^{n}\omega_{i}x_{i} -\mathcal{C}\right)^2,
\end{equation}
where $\mathcal{A}$ and $\mathcal{B}$ are adjustable coefficients. This function reaches its minimum when $ \sum_{i=1}^{n} \omega_{i} x_{i} = \mathcal{C} -\mathcal{A}/2\mathcal{B}$. By tuning  $ A $ and $ B $ appropriately, we capture all possible weight sums less than $ \mathcal{C} $, identifying all feasible partial solutions (FPS). Subsequently, we will use ``FPS" as the solutions where the total weight of items in a bin that does not exceed the upper bound $\mathcal{C}$. 
To adjust the item weights, we use  $ \mathcal{A}= 2\mathcal{B}(\mathcal{C} - k\Delta\omega)$, where $ \Delta\omega $ represents the minimum difference between any two item weights. The parameter $ k $ varies from 1 to $ \mathcal{C}/\Delta\omega $, enabling the desired weight sum to fall within the range $ [\Delta\omega,~\mathcal{C}] $. Consequently, when $ \mathcal{B} $ is fixed, $ \mathcal{A} $ becomes a variable dependent on $ k $. However, determining an appropriate step size for $ k $ is not trivial. If $ k $ is too large, the method may fail to capture enough solutions. Conversely, if $k$ is too small, the inclusion of an excessive number of sub-Hamiltonians could lead to a significant increase in computational resources. To address this trade-off, the $ k $-stepsize is treated as a tunable variable in Sec.~\ref{sec: results}.

To map Eq.~\eqref{eq:h_p_2_mod} onto the qubit basis, we utilize the relation $ X_{i} = (I - \sigma^{i}_{z})/2 $, where $I $ is the identity operator and $ \sigma^{i}_{z} $ represents the Pauli-Z operator acting on qubit $ i $. After this transformation, the resulting Hamiltonian is given by:
\begin{align}
\label{eq:H_P_mod}
    H_c = \sum_{i=1}^{n}\left[-\frac{\mathcal{A}}{2} + \mathcal{B}\left(\mathcal{C} - \frac{1}{2}\sum_{i=1}^{n}\omega_{i}\right)\right]\omega_{i}\sigma^{i}_{z} + \sum_{i<j}\frac{\mathcal{B}}{2}\omega_{i}\omega_{j}\sigma^{i}_{z}\sigma^{j}_{z}.
\end{align}
The next step is to determine the ground state of this set of cost Hamiltonians using quantum algorithms. While quantum annealing algorithms have been proposed \cite{de2022hybrid,v2023hybrid}, our focus will be on the application of VQAs. In particular, hereinafter we will use QAOA assisted by CD driving.

\section{Quantum Sampling algorithms for 1dBPP}
\label{sec: algorithm}

\begin{figure}[]
    \centering
    \includegraphics[width=\linewidth]{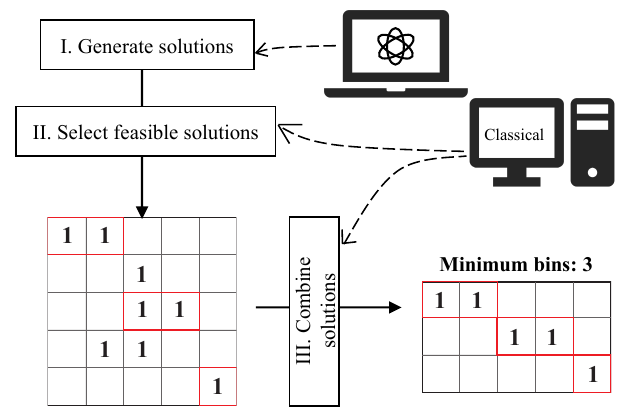}
    \caption{Schematic of the procedure proposed for solving 1dBPP. Step I: Starting from an empty bin, quantum circuits are applied to obtain partial solutions corresponding to several weight sums. Step II: A  feasible partial solutions $\mathcal{S}$ set is obtained by classically selecting. Step III: Using a classical process, we combine the feasible partial solution to pack all the items.}
    \label{fig: method illu}
\end{figure}

As illustrated in Fig.~\ref{fig: method illu}, the method consists of three steps. Step I involves a parameterized quantum circuit, as described in Algorithm~\ref{tab: algo}, where partial solutions are obtained from the ground states of the subset Hamiltonians. We traverse the $k$ value to derive the Hamiltonian over the entire interval, and then apply the QAOA algorithm and CD ansatz, instead of quantum annealing, to compute the ground state energy and obtain a valid subset of solutions. Step II acts as a filter, as some partial solutions generated in Step I may be infeasible partial solutions (IPS), where the weight sum exceeds the bin capacity $\mathcal{C} $. A simple classical check is performed to determine whether each partial solution violates the capacity constraint.  Step III combines FPS using multiple bins to pack the items. The process begins with a single bin and continues until all items are successfully packed. The final solutions (FS) for 1dBPP are thus obtained.

\begin{algorithm}[!tb]
\SetAlgoLined
\KwIn{$w_i$, $\mathcal{C}$, $\mathcal{B}$, $\mathcal{A}$}
\KwOut{The feasible partial solution set $\mathcal{S}$}
\For{$k = 1,\ldots,\mathcal{C}/\Delta_w$}{
Update Hamiltonian parameter, $\mathcal{A} = 2\mathcal{B}(\mathcal{C} - k\Delta_w)$\;
Run the parameterized circuit to obtain FPS set $\mathcal{S^{*}}$ corresponding to a single weight sum\;
Add the result to the set, $\mathcal{S} \gets \mathcal{S^{*}}$\;
}
\caption{{\sc Subset sampling routine}}
\label{tab: algo}
\end{algorithm}

\begin{figure*}[!t]
    \centering
    \includegraphics[width=0.9\linewidth]{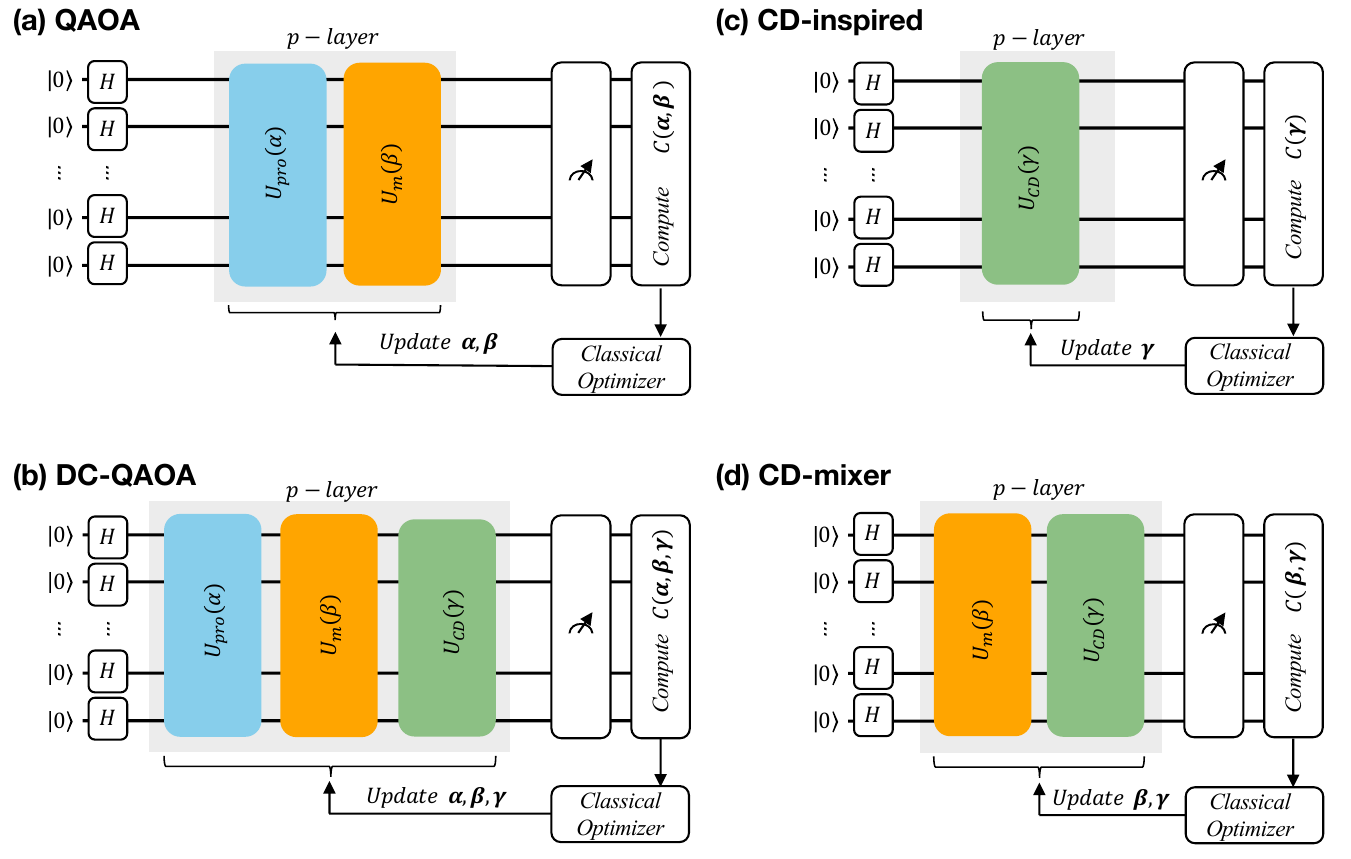}
    \caption{The quantum circuit structures of QAOA in (a), DC-QAOA in (b), CD-inspired ansatz in (c), and CD-mixer in (d) are illustrated. Starting from the $\ket{0}$ states, Hadamard gates prepare an equal superposition of the entire computational basis, which contains the optimal solutions. Then, four ansatz operations are applied to the states for $p$ layers, respectively. The notation $\langle C \rangle$ denotes the cost function of each ansatz, which will be optimized by classical optimizers. The variational circuit will identify the set of optimal parameters and feed them back into the corresponding circuit to sample the states.
}
    \label{fig:cir}
\end{figure*}

\subsection{QAOA}
QAOA is a quantum algorithm specially designed to tackle optimization problems by leveraging parameterized quantum circuits~\cite{farhi2014quantum}, aiming to enhance the efficiency of quantum adiabatic algorithms. As illustrated in Fig. \ref{fig:cir}(a), the evolution operator $U(\bm{\alpha},\bm{\beta})$ is constructed by alternately applying mixer and Hamiltonian operators:
\begin{equation}
    U(\bm{\alpha},\bm{\beta}) = \prod_{j=1}^{p}e^{-i\beta_{j}H_{m}} e^{-i\alpha_{j}H_{c}},
\end{equation}
where $H_{m}$ and  $H_c$ represent the mixer term and cost Hamiltonian, respectively. The parameters $\bm{\alpha} = \{\alpha_{1},\dots,\alpha_{p}\}$ and $\bm{\beta} = \{\beta_{1},\dots,\beta_{p}\}$ correspond to $p$ layers of circuits and are optimized to minimize the cost function. The mixer Hamiltonian facilitates the preparation of an easily prepared initial state, while the cost Hamiltonian $H_c$ encodes the specific optimization problem, defining the energy landscape over which the algorithm searches for the minimum.

To implement QAOA, we start by defining a cost function  expressed in terms of the parameters embedded in the quantum circuit:
\begin{equation}
    C(\bm{\alpha},\bm{\beta}) = \bra{\psi(\bm{\alpha},\bm{\beta})}H_c\ket{\psi(\bm{\alpha},\bm{\beta})}.
\end{equation}
The state $\ket{\psi(\bm{\alpha}, \bm{\beta})}$ is obtained by applying the parameterized gates in an alternating fashion to the initial state 
\begin{equation}
    \ket{\psi(\bm{\alpha}, \bm{\beta})} = U(\bm{\alpha}, \bm{\beta}) \ket{+}^{\otimes n},
\end{equation}
where $\ket{+}^{\otimes n} = [(\ket{0} + \ket{1})/\sqrt{2}]^{\otimes n}$ is the equal superposition over $n$ qubits.

In the context of the 1dBPP, the cost Hamiltonian $H_c$ in Eq.~\eqref{eq:H_P_mod} is chosen to encode the constraints and objectives of the problem, specifically targeting the minimization of the number of required bins. The mixer term is defined as $H_{m} = \sum_{i=1}^{n}\sigma^{i}_{x}$. Given the similarity between 1dBPP and other combinatorial problems like MaxCut, traveling salesman problem (TSP)~\cite{amaro2022case,montanez2024transfer}, this choice of mixer term has proven effective in our conducted simulations and experiments.

\subsection{DC-QAOA and its variants}

\subsubsection{DC-QAOA} 

To begin, we introduce the DC-QAOA, a hybrid quantum-classical algorithm that extends the standard QAOA by incorporating CD terms into the quantum circuit, see Fig. \ref{fig:cir}(b). This modification facilitates faster convergence and enhances the quality of solutions for combinatorial optimization problems~\cite{PhysRevResearch.4.013141}. The evolution operator $U(\bm{\alpha},\bm{\beta}, \bm{\gamma})$ introduces an additional set of parameters $\bm{\gamma}$:
\begin{equation}
    U(\bm{\alpha},\bm{\beta}, \bm{\gamma}) = \prod_{j=1}^{p}e^{-i\beta_{j}H_{m}} e^{-i\alpha_{j}H_{c}}  e^{-i\gamma_{j}H_{CD}},
\end{equation}
where $H_{CD}$ represents the CD terms. However, computing exact CD terms can be computationally expensive. To address this, an approximate method known as the nested commutator (NC) approach can be employed, providing a tractable approximation of the CD terms~\cite{PhysRevLett.123.090602,PhysRevApplied.15.024038,PhysRevResearch.6.013147}. Specifically, the NC is given by: 
\begin{equation}
\label{gauge}
\hat{A}_\lambda^{(l)} = i\sum_{k=1}^l\alpha_k(t)\mathop{\underbrace{[\hat{H}_a, [\hat{H}_a,\cdots[\hat{H}_a}}\limits_{2k-1}, ~ \partial_\lambda \hat{H}_{a}]]],
\end{equation}
where $l$ is the expansion order, and $\hat{H}_{a}= (1-\lambda) H_{m} + \lambda H_{c}$ defines the adiabatic interpolation between the mixer and cost Hamiltonians, with the time-dependent scheduling function varying from $0$ to $1$. This method constructs approximate CD terms that effectively suppress non-adiabatic transitions, enabling efficient quantum evolution even with constrained computational resources~\cite{PhysRevLett.123.090602}. In our approach for 1DBPP, we restrict ourselves to one-body and two-body interactions, and write down the explicit form of CD Hamiltonian as follow:
\begin{equation}
\label{CD}
    H_{CD}=\sum_{ij} (\sigma_y^i\sigma_z^j + \sigma_z^i\sigma_y^j) + \sum_{i=1}^n \sigma_y^i,
\end{equation}
through the gauge potential $\hat{A}_\lambda^{(l)}$, see Eq. (\ref{gauge}). 

In the optimization process, the difference between QAOA and DC-QAOA lies in the incorporation of additional parameters for the CD terms in each circuit layer in the latter. As a result, the cost function is reformulated as
\begin{equation}
\label{cost}
    C(\bm{\alpha},\bm{\beta}, \bm{\gamma}) = \bra{\psi(\bm{\alpha},\bm{\beta}, \bm{\gamma})}H_{c}\ket{\psi(\bm{\alpha},\bm{\beta}, \bm{\gamma})},
\end{equation}
where $\bm{\gamma} = \{\gamma_{1},\dots,\gamma_{p}\}$ represents the parameters associated with the $p$-layer unitary evolution operator $U_{CD}( \bm{\gamma})$, selected from a predefined CD Hamiltonian Eq.~\eqref{CD}. Although this approach increases the number of parameters to be optimized per layer, DC-QAOA compensates by reducing the required number of layers to achieve convergence. By leveraging this trade-off, DC-QAOA effectively optimizes complex combinatorial problems such as the 1dBPP on near-term quantum hardware, striking a balance between circuit depth and solution accuracy.

\subsubsection{CD-inspired ansatz} 
Rather than combining CD terms with the original QAOA, one can employ a simplified CD-only ansatz in Fig. \ref{fig:cir}(c), which reduces circuit complexity while retaining the ability to accelerate quantum evolution~\cite{PhysRevApplied.20.014024, romero2024optimizing, tang2024exploring, cadavid2024bias, romero2024biasfielddigitizedcounterdiabaticquantum}.
The unitary evolution operator for this CD-inspired ansatz, which depends solely on the parameter set $ \bm{\gamma} $, is expressed as
\begin{equation}
    U(\bm{\gamma}) = \prod_{j=1}^{p} e^{-i\gamma_{j}H_{CD}}.
\end{equation}
In this simplified framework, the cost function Eq.~\eqref{cost} reduces to 
$C(\bm{\gamma})$,  further streamlining the optimization process.
This ansatz has been shown to yield reliable estimates of the ground state while substantially reducing both the number of quantum gates required and the optimization overhead associated with parameter tuning. These advantages make the CD-inspired ansatz particularly suitable for solving the 1dBPP on near-term quantum hardware, where minimizing circuit depth and parameter complexity is critical for practical implementations.

\subsubsection{CD-mixer ansatz}  

In addition to the standard DC-QAOA and CD-inspired variant, we also draw inspiration from the ADAPT-QAOA~\cite{zhu2022adaptive}. Our goal is to incorporate a simple mixer Hamiltonian into the CD-inspired ansatz. This adjustment may allow the CD-inspired ansatz to initiate the optimization process in a region closer to the global minimum without excessively expanding the search space:
\begin{equation}
    U(\bm{\beta}, \bm{\gamma}) = \prod_{j=1}^{p}e^{-i\beta_{j}H_{m}}e^{-i\gamma_{j}H_{CD}}.
\end{equation}
With this formulation, the cost function Eq.~\eqref{cost} simplifies to
$C(\bm{\beta},\bm{\gamma})$, keeping two sets of parameters while maintaining effective quantum evolution.
Since the mixer Hamiltonian only requires a single $ \sigma_x $ operator per qubit, the resulting circuit remains significantly shorter than standard QAOA and is straightforward to implement. A detailed schematic of the proposed ansatz is shown in Fig.~\ref{fig:cir}(d).

\begin{figure}[!t]
    \centering
    \includegraphics[width=\linewidth]{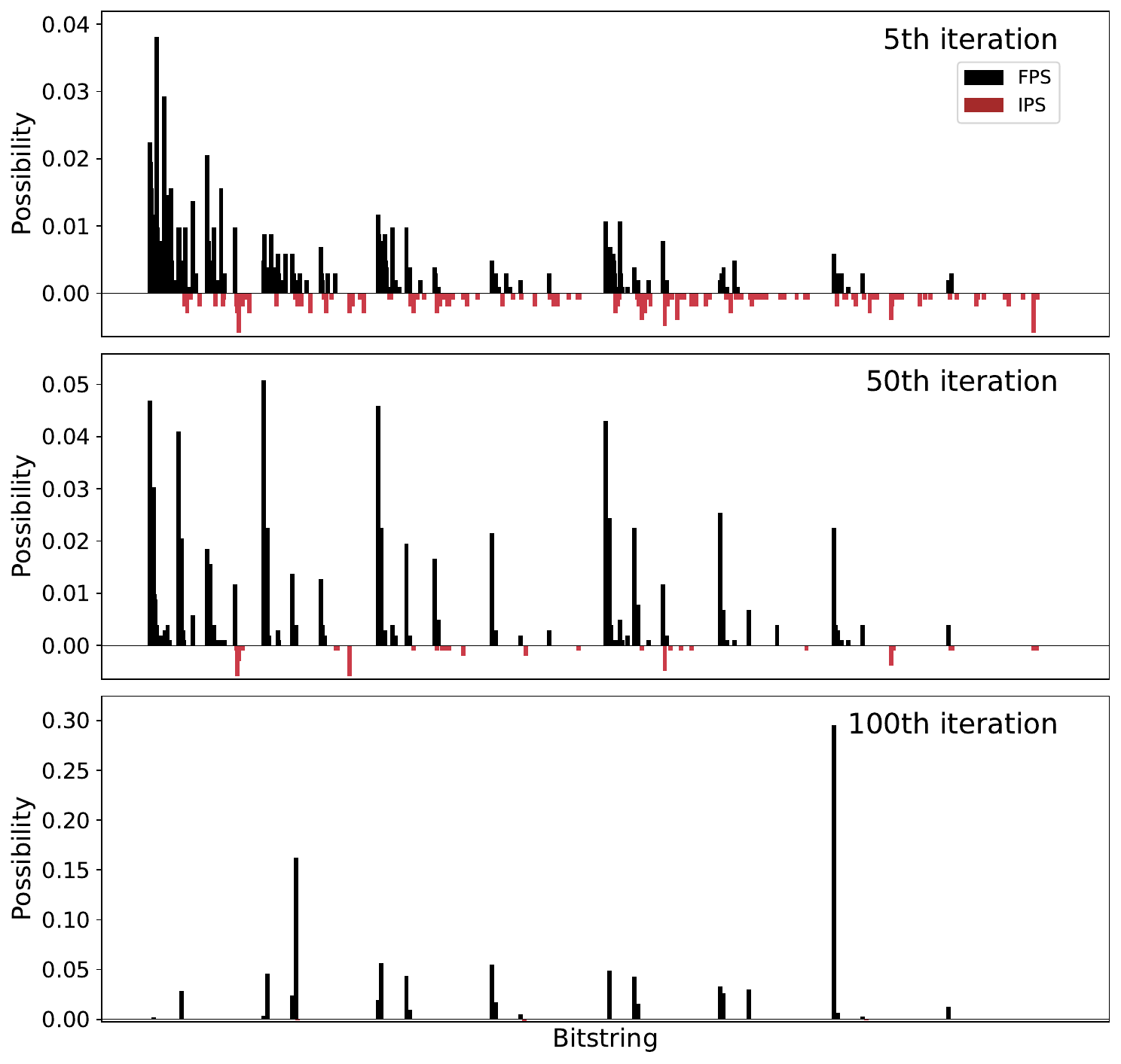}
    \caption{State probabilities at the 5th, 50th, and 100th iterations using a one-layer CD-mixer ansatz with $\mathcal{A}/2\mathcal{B} = 60$. Positive values represent configurations where the total weight is within the bin capacity, while negative values indicate configurations exceeding the bin capacity. The results highlight the impact of iterations on the solution quality, affecting both the number of partial solutions and their probability distribution.
}
    \label{fig: ite-bitstring}
\end{figure}%

\section{Numerical Simulations}
\label{sec: results}

We perform numerical simulations using various sets of weight values as inputs to quantum circuits to solve the 1dBPP. The bin capacity is set to be $\mathcal{C}=120$, while the item weights are randomly generated following a real-world-oriented distribution~\cite{v2023hybrid, doi:10.1287/ijoo.2019.0013, OSABA2023109309} 
within different weight ranges. 
This capacity value can be chosen arbitrarily, as weights are generated accordingly to reflect realistic scenarios, enhancing the generality and applicability of our quantum algorithms.

First, we examine how the optimization process generates and selects partial solutions. As shown in Fig.~\ref{fig: ite-bitstring}, under a specific $\mathcal{A}/2\mathcal{B}$, which defines a particular capacity in this context, many IPS appear within the first five iterations. As the optimization progresses, the system gradually converges, eliminating IPS by the 100th iteration. The presence of multiple near-ground-state solutions, satisfying the constraints in Eq.~\eqref{eq:h_p_2_mod}, is attributed to small weight differences that make distinguishing these states more challenging.

To further analyze performance, we investigate the impact of different item weights, by focusing on two key variables: the \( k \)-stepsize and the number of iterations. Both parameters influence the balance between the computing resources and solution quality. To mitigate the effects of random initialization, we performed five independent trials for each instance.

In order to isolate variable effects, we vary one parameter while keeping the other fixed. Specifically, the $k-$stepsize is adjusted within the range $[1, 4]$ in increments of $1$, while the number of iterations is varied from $20$ to $100$ increments of $20$. We also introduce variations in the weight standard deviations to assess performance consistency. Regarding this, the Adam optimizer in MindSpore Quantum frame~\cite{mindquantum} is employed due to its reliable convergence within relatively few iterations.  

\begin{figure*}[t]
    \centering
    \includegraphics[width=\linewidth]{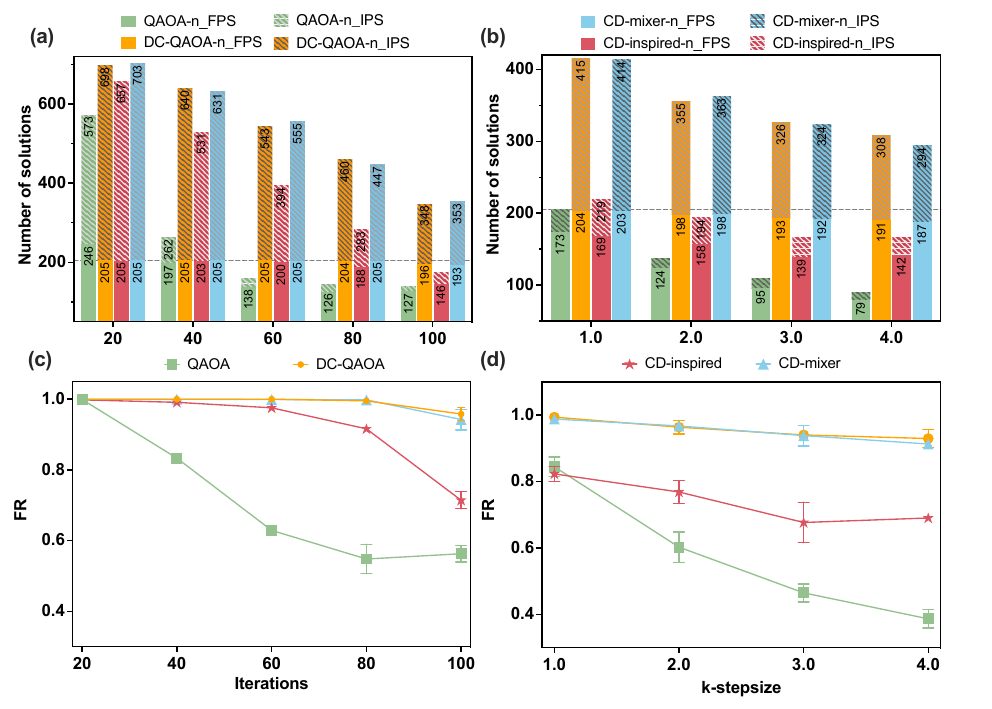}
    \caption{(a)-(d) Number of partial solutions and FR in Eq.~\eqref{eq:AR} as a function of k-stepsize and iterations for the instance W3\_10. Four ansatz including QAOA (green), DC-QAOA (orange), CD-inspired (red), and CD-mixer (blue) are tested and compared. (a)-(b): The pattern bars are the number of infeasible partial solutions (IPS) and colored bars are feasible partial solutions (FPS). The grey dashed line is the number of exact solutions. (c)-(d): After considering five random initializations of the circuit parameters, the error bars indicate the uncertainty associated.}\label{fig:W3}
\end{figure*}%

To benchmark the performance, we define the following metrics:
\begin{itemize}
    \item  \textit{Minimum number of bins:} The minimum bin count required to pack all items, serves as a fundamental performance indicator.
    \item \textit{Feasibility ratio:} The proportion of
    FPS within a single bin generated by the algorithm: 
      \begin{equation}\label{eq:AR}
        \text{FR} = \frac{\text{number of FPS}}{\text{number of exact partial solutions}}.
    \end{equation}
   Here the exact number of partial solutions is obtained through brute force.  An FR value of 1 indicates full coverage of feasible solutions.
    \item \textit{Final solution numbers:} The total number of FPS and IPS provided by the algorithm.
\end{itemize}

We compare four ansatz schemes based on these metrics across different iteration counts and $k$-stepsize. In Figs.~\ref{fig:W3}(a)-(d), the gray dashed line represents the exact number of partial solutions, while the green, orange, red, and blue bars correspond to the FPS results from QAOA, DC-QAOA, CD-inspired ansatz, and CD-mixer ansatz, respectively. Patterned bars represent IPS identified from the quantum approach.

\begin{table*}[!t]
\centering
\caption{Instances information generated from the real-instance-based dataset~\cite{v2023hybrid}, including the standard derivative (std) of item weights, the theoretical optimal number of bins, and the corresponding exact number of FS for each instance. The best performance of all ansatz schemes is listed accordingly.}
\begin{ruledtabular}
\begin{tabular}{lccccccc}
Instance &  Std / weight range  & \# of items & \makecell{ Optimal bins / \\ \# of final solutions}  & QAOA & DC-QAOA & CD-inspired & CD-mixer\\ \midrule
W1\_10   &  10.4 / [40,69] &    10       & 5 / 100800 & 5 / 540  & 5 / 805 & 5 / 840  & \textbf{5 / 840} \\
W2\_10   &  9.4 / [20, 52]  &    10       & 3 / 3708   & 3 / 618  & 3 / 618 &   3 / 612  & \textbf{3 / 618} \\
W3\_10  &  7.9 / [21, 49]  &    10       & 3 / 1368   & 3 / 228  & 3 / 228 & 3 / 228  & \textbf{3 / 228}\\
W4\_10   &  6.3 / [21, 42]  &     10       & 3 / 6168   & 3 / 1016 & 3 / 1028& 3 / 1017 & \textbf{3 / 1028}\\
W5\_10   &  4.4 / [42, 80]  &     10       & 7 / 378000 & 7 / 75   & 7 / 75  & 7 / 75   & \textbf{7 / 75}\\
W1\_12   &  9.5 / [21, 54]  &     12       & 4 / 213024 & 4 / 8774 & \textbf{4 / 8828}   & 4 / 8805  & 4 / 8805\\
W2\_12   &  4.9 / [20, 36]  &     12       & 3 / 33684 & 3 / 5117   & 3 / 5360  & 3 / 5339   & \textbf{3 / 5590}
\end{tabular}
\label{tab: num solution}
\end{ruledtabular}
\end{table*}

Our objective is to find all FPS and filter out as many IPS as possible by optimizing the circuit parameters. As seen in Fig.~\ref{fig:W3}(a),  when the circuit layer is set to 
$p=1$ and the $k-$stepsize is $2.5$, insufficient iterations result in numerous infeasible solutions. However, as the number of optimization iterations increases, the number of infeasible solutions generated by the four ansatz schemes gradually decreases. Therefore, an optimal number of iterations can be identified to produce a sufficient number of feasible solutions without requiring excessive computational resources. However, the optimal number of iterations varies across the four ansatz schemes, depending on the specific way the mixer, cost, and CD Hamiltonians are combined or selected. In detail, QAOA can filter out infeasible solutions with fewer iterations, whereas the CD-inspired ansatz requires more iterations. Additionally, as shown in Fig.~\ref{fig:W3}(c), QAOA exhibits high sensitivity to the number of iterations, resulting in a loss of feasible partial solutions as the iteration count increases. This sensitivity poses a challenge in the subsequent step, where partial solutions are combined to obtain final solutions. In contrast,  DC-QAOA and CD-mixer maintain consistently high FR values near 1, demonstrating superior robustness and efficiency in FPS generation.

The choice of $k$-stepsize significantly impacts quantum resource utilization, making it crucial to balance the step size and resource efficiency. A smaller step size generates more sub-Hamiltonians, thereby increasing quantum measurement demands. Consequently, we aim to obtain sufficient solutions by using a larger $k$-stepsize. As shown in Fig.~\ref{fig:W3}(b), increasing the step size causes QAOA to lose a substantial number of partial solutions, while the CD-inspired ansatz continues to produce all feasible solutions, albeit with some infeasible ones still present. Thus, CD-inspired ansatz outperforms conventional QAOA by reducing infeasible solutions and demonstrating better compatibility with the capabilities of current quantum hardware.
Focusing on the FR shown in Fig.~\ref{fig:W3}(d), CD-inspired ansatz significantly outperforms usual QAOA as the $k$-stepsize increases. In particular, DC-QAOA and CD-mixer ansatz maintain a high FR even at a $k$-stepsize of 4, requiring only 30 quantum circuit measurements. By contrast, QAOA achieves a high FR only when the $k$-stepsize is small, e.g. $0.5$, which demands nearly $240$ measurements--a substantially more time-consuming and computationally intensive process. Another variable with importance is the number of layers of circuits, it doesn't affect the results that much. The analysis is detailed in Appendix~\ref{sec: layer}.

This analysis highlights the efficiency of DC-QAOA and its different variants (CD-inspired ansatz and CD-mixer ansatz) to effectively explore the partial solution space, making them well-suited for near-term quantum computing applications. The details of these instances, along with the number of final solutions obtained from each ansatz, are provided in Table~\ref{tab: num solution}. By varying the iterations, and $k$-stepsize, we figure out the performance for each ansatz and record the results in this table. When compared to the exact final solutions and corresponding bin counts, all ansatz schemes correctly predict the number of bins while generating significantly fewer solutions than brute force methods. Through the evaluation of various weight distributions, we observe consistent performance across all ansatz models. However, we also find that the quantum protocol encounters challenges when more items are packed into a single bin, due to the degeneracy of eigenstates. Conversely, this issue is less pronounced when fewer items are packed. Although the hybrid quantum-classical approach produces fewer solutions than brute force methods, it demonstrates strong potential as a practical tool, particularly as quantum computing capabilities continue to advance rapidly.

\section{Implementation on IBM Quantum Computer}
\label{sec: exp}

A comparative summary of quantum gate counts for different ansatz formulations is presented in Table~\ref{tab: gate number}. From a simulation perspective, the CD-inspired ansatz employs a shallower circuit, enabling it to achieve superior performance by minimizing the number of CNOT gates and reducing the number of parameters to optimize. This decrease in gate complexity and parameter overhead enhances the feasibility and efficiency of its implementation on current quantum hardware.  

\begin{table}[!t]
\centering
\caption{The number of quantum gates required per layer in the simulation is summarized, where $n$ denotes the number of items (qubits). The counts of CNOT gates and parameterized gates serve as key indicators for assessing the practicality of each ansatz.}
\begin{ruledtabular}
\begin{tabular}{lccc}
Ansatz & Parameterized gates & CNOT gates & Total gates \\
\midrule
QAOA & $\binom{n}{2} + n$ & 2$\binom{n}{2}$ & $3\binom{n}{2} + 2n$\\
DC-QAOA & $2\binom{n}{2}$ & 4$\binom{n}{2}$ & $8\binom{n}{2} + 3n$\\
CD-inspired & $\binom{n}{2}$ & 2$\binom{n}{2}$ & $5\binom{n}{2} + 2n$\\
CD-mixer & $\binom{n}{2} + n$ & 2$\binom{n}{2}$ & $5\binom{n}{2} + 3n$\\
\end{tabular}
\label{tab: gate number}
\end{ruledtabular}
\end{table}

To further demonstrate the effectiveness of the CD terms, we execute the quantum circuit on IBM quantum computer. The circuits illustrated in Figs.~\ref{fig:cir}(a)-(d) represent the initial designs, which were first tested via classical simulations before deployment on real devices.
Implementation on IBM quantum computer is crucial, given its specific native gates, which necessitate effective transpilation due to its reliance on specific native gate sets and its topology constraints defined by the qubit connectivity map. The circuits discussed in Sec.~\ref{sec: results} are constructed without consideration of the actual layout or hardware-specific error characteristics. Therefore, it is essential to adapt these circuits for real-device execution by transpiling them into hardware-adaptive gate configurations. Transpilation involves several steps: first, converting high-level gates into the IBM-native gate set; second, optimizing the circuit to minimize depth, gate count, and especially CNOT operations, which are typically more error-prone than single-qubit gates.

\begin{figure}[!t]
    \centering
    \includegraphics[width=\linewidth]{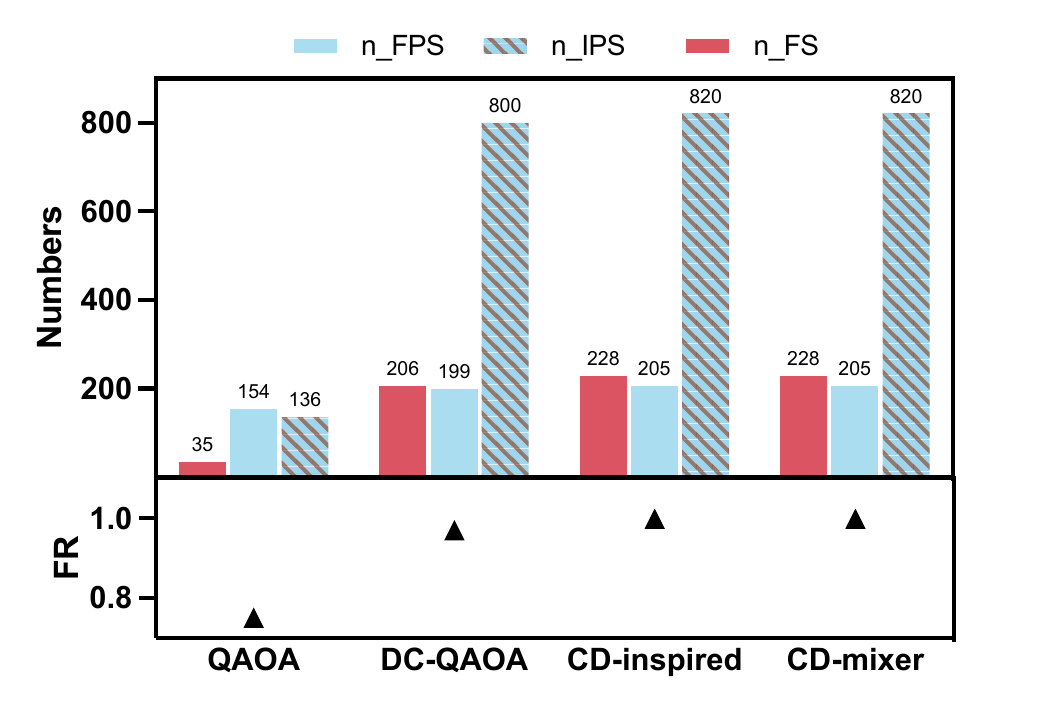}
    \caption{Experimental results for instance W3\_10. The upper part of the figure displays the number of FPS (solid blue bar) and IPS (pattern blue bar) for a single bin. Additionally, the number of FS for the 1dBPP problem is represented by the red columns. The lower part of the figure illustrates the FR. Parameters are set as follows: number of iterations is 100, $k-$stepsize of 4, and single ansatz layer. }
    \label{fig:exp}
\end{figure}%

Furthermore, the qubit assignment is carefully selected based on the calibration data provided by IBM, prioritizing qubits with higher coherence times and lower readout errors. Noise-aware optimization techniques, such as dynamical decoupling and readout error mitigation, are applied when possible to improve the reliability of the measurement outcomes. This systematic preparation aims to reduce the adverse effects of decoherence, crosstalk, and gate errors in the experiments, ensuring more accurate results when comparing the performance of different ansatz schemes. The experiments are conducted on IBM Quantum devices using the Qiskit framework, which provides essential tools for circuit transpilation, noise modeling, and execution. Multiple runs have been performed for each configuration to account for statistical fluctuations and device noise, with results analyzed to evaluate the feasibility and efficiency of QAOA, DC-QAOA and its variants under real hardware conditions. 

Next, we select the variable configurations that demonstrated the best performance during the simulation phase and replicate these conditions on the 127-qubit IBM quantum platform \textsc{ibm\_strasbourg} \cite{IBM}, whose layout is provided in Appendix~\ref{sec: IBM detail}. For this experiment, the chosen parameters are $k = 4$, $100$ iterations, and a single layer. It is important to emphasize that the optimization process in this context is highly sensitive to the choice of initial parameters and the gradient calculation method, reflecting the intricate structure of the cost landscape. To address this, we use the parameter shift rule for gradient estimation, combined with the Adam optimizer~\cite{kingma2014adam} for parameter updates, using a learning rate set at $0.05$. As illustrated in Fig.~\ref{fig:exp}, DC-QAOA and its variants achieve an almost $100\%$ match with the exact solutions for single-bin packing, while previous QAOA attains only $75\%$. This discrepancy directly affects the number of final solutions generated for the 1dBPP, as highlighted by the red columns. The results in the upper segment correspond to the simulated outcomes, where the enhanced QAOA with CD ansatz consistently produces more IPS while maintaining a high FR. One contributing factor to this superior performance lies in the structure. As presented in Table~\ref{tab: exp gate number}, the CD-mixer ansatz offers greater expressibility than QAOA due to the distinct nature of its mixer Hamiltonian, which facilitates more efficient exploration of the solution space. While all tested ansatz formulations successfully identify the optimal bin number $ M=3 $, QAOA yields fewer final solutions overall, underscoring the practical advantage of integrating CD  terms into the quantum optimization framework.

\begin{table}[t]
\centering
\caption{The number of required native quantum gates per layer for \textsc{ibm\_strasbourg}. Instance W3\_10 is examined using the IBM transpilation technique, so the quantities may not strictly align with the values in Table~\ref{tab: gate number}.}
\begin{ruledtabular}
\begin{tabular}{lcccc}
Ansatz & Depth & CNOT gate & ECR gate & Total gates \\
\midrule
QAOA & 61 & 90 & 161 & 225\\
DC-QAOA & 96 & 180 & 367 & 432\\
CD-inspired & 59 & 90 & 168 & 228\\
CD-mixer & 61 & 90 & 168 & 247\\
\end{tabular}
\label{tab: exp gate number}
\end{ruledtabular}
\end{table}%

\section{discussion and conclusion}
\label{sec: conclusion}
The 1dBPP, as an NP-hard problem, has historically received less attention compared to other combinatorial optimization tasks, primarily due to its exponential growth in complexity as the number of items to be packed increases. The inherent constraints in the original Hamiltonian are particularly difficult to encode onto present-day quantum devices with limited qubit connectivity and coherence times. To address this, we simplified the Hamiltonian by focusing on single-bin packing and integrating it with a classical algorithm, allowing for a more efficient quantum-classical hybrid approach. Our study has demonstrated that the variational quantum ansatz, particularly CD-inspired ansatz and CD-mixer ansatz, offers promising results for solving instances with realistic weight distributions, making it viable for implementation on real quantum hardware.

A key takeaway from this study is the comparative performance of conventional QAOA, DC-QAOA and its variants, analyzed both in simulations and on IBM quantum computer. The integration of CD terms with distinct combinations of cost and mixer Hamiltonians has shown a clear advantage in accelerating the system’s convergence to the target state and enhancing stability despite fluctuations in circuit parameters. The CD-mixer ansatz, in particular, significantly maintains solution accuracy, albeit at the cost of an increased parameter space. This trade-off highlights the necessity of optimizing circuit training procedures to enable effective scaling of the method to larger instances of BPP. Additionally, the concept of Hamiltonian simplification could be extrapolated to other complex combinatorial problems, offering a potential pathway for adapting quantum algorithms to tackle optimization tasks that are otherwise infeasible on current quantum hardware. This study suggests that carefully designed ansatz modifications, such as the combination of CD terms with the cost and/or mixer Hamiltonian, can play a crucial role in improving quantum algorithms for combinatorial optimization.

Despite the observed improvements, several open questions remain. One crucial aspect is understanding the relationship between the CD-enhanced solution space and the standard QAOA framework. While the CD ansatz schemes tend to generate a higher number of IPS, the underlying mechanism remains uncertain. A deeper theoretical analysis is required to determine whether this phenomenon stems from a fundamental change in the search space or from algorithmic constraints.
In the context of ansatz schemes, the analysis of the commutator terms in the first-order Trotter-Suzuki product expansion reveals that QAOA has a limited search space with small Trotter errors (16 terms), but its optimization is challenging due to parameter selection. DC-QAOA expands the search space with additional CD terms, increasing Trotter errors with 61 terms while improving the optimization landscape. The CD-mixer scheme, on the other hand, modifies the evolution path without expanding the search space, enhancing expressivity and potentially reducing errors (with 16 different terms) for more stable convergence. This raises important questions about the trade-off between expressivity, error accumulation, and convergence speed in ansatz design.
Potential avenues for exploration include the Lie algebraic theory of barren plateaus \cite{ragone2024lie} and the role of noncommuting Hamiltonians sharing their eigenstates \cite{PhysRevResearch.6.023243}.
Additionally, the cost function landscape in variational quantum algorithms remains a significant challenge. The rugged nature of these landscapes makes optimization highly sensitive to initialization strategies and choice of optimizer. Further research is needed to identify which optimization techniques—such as second-order methods \cite{secondorder} or machine-learning-assisted optimizers \cite{kundu2024reinforcement, sciorilli2025towards}—are best suited for navigating these landscapes effectively

Last but not least, to extend this approach to larger bin packing problems~\cite{LODI2002379,v2023hybrid} with more complex constraints is a natural progression, which remains to be future work. Investigating hybrid quantum-classical strategies that leverage quantum variational techniques in conjunction with advanced classical optimization methods could further enhance scalability and performance. Finally, understanding the interplay between cost function complexity and ansatz structure may lead to new insights for designing more efficient and robust quantum circuits applicable to a broader range of combinatorial optimization problems beyond 1dBPP \cite{romero2024biasfielddigitizedcounterdiabaticquantum,copovergraph}.

\begin{figure}
    \centering
    \includegraphics[width=\linewidth]{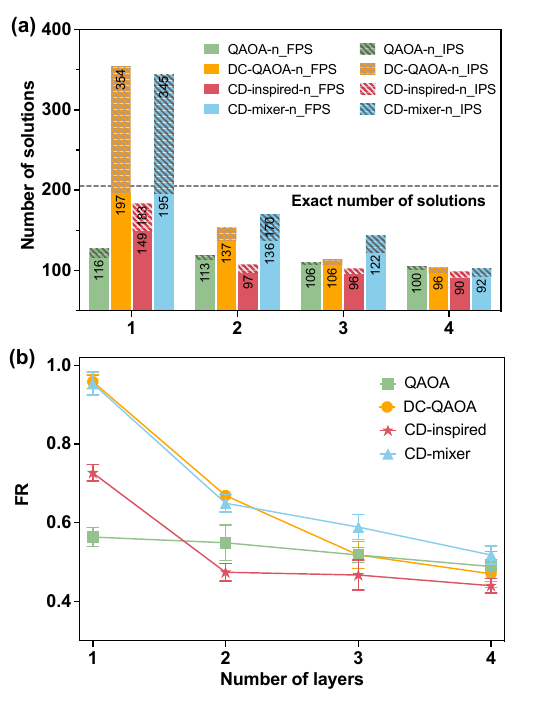}
    \caption{The number of solutions (a) and FR (b) are plotted as functions of the number of layers. Four ansatz schemes—QAOA (green), DC-QAOA (orange), CD-inspired (red), and CD-mixer (blue)—are tested and compared. The patterned bars represent the number of infeasible solutions, while the gray dashed line indicates the number of exact solutions. Error bars, shown as vertical lines on the symbols, illustrate the variability in the results.} \label{fig: layer}
\end{figure}%

\section*{Availability of data and materials}
The code and data, as well as all the results discussed in this work, are available at \url{https://github.com/Ruoqian-Xu/Digitized-Counterdiabatic-Optimization-Algorithms-for-Bin-Packing-Problem}. \\\\

\section*{Competing interests}
The authors declare that they have no competing interests.

\section*{Authors' contributions}
X.C. and Y.B. conceived the research and designed the study. R.X. and S.V.R. performed the theoretical analysis and simulations. R.X. S.V.R. and J.T. contributed to the interpretation of the results and provided critical insights. All authors discussed the results, contributed to the writing, and approved the final manuscript.
 
\begin{acknowledgments}
This work is partially supported by the Project Grant No. PID2021-126273NB-I00 funded by MCIN/AEI/10.13039/501100011033 and by “ERDF A way of making Europe” and “ERDF Invest in your Future”, 
the Spanish Ministry of Economic Affairs and Digital Transformation through the QUANTUM ENIA project call-Quantum Spain project, and CPS-MindSpore funding. 
R.Q.X. and J.L.T. appreciate the support from China Scholarship Council (CSC) under Grant Nos: 202206890001 and 202306890004. 
Y.B. acknowledges ayudas Ramón y Cajal (RYC2023-042699-I). We acknowledge the use of IBM Quantum services for this work. The views expressed are those of the authors and do not reflect the official policy or position of IBM or the IBM Quantum team.
\end{acknowledgments}

\begin{appendices}
\appendix
\section{Detailed study on circuit layers} 
\label{sec: layer}

\begin{figure}[t]
    \includegraphics[width=\linewidth]{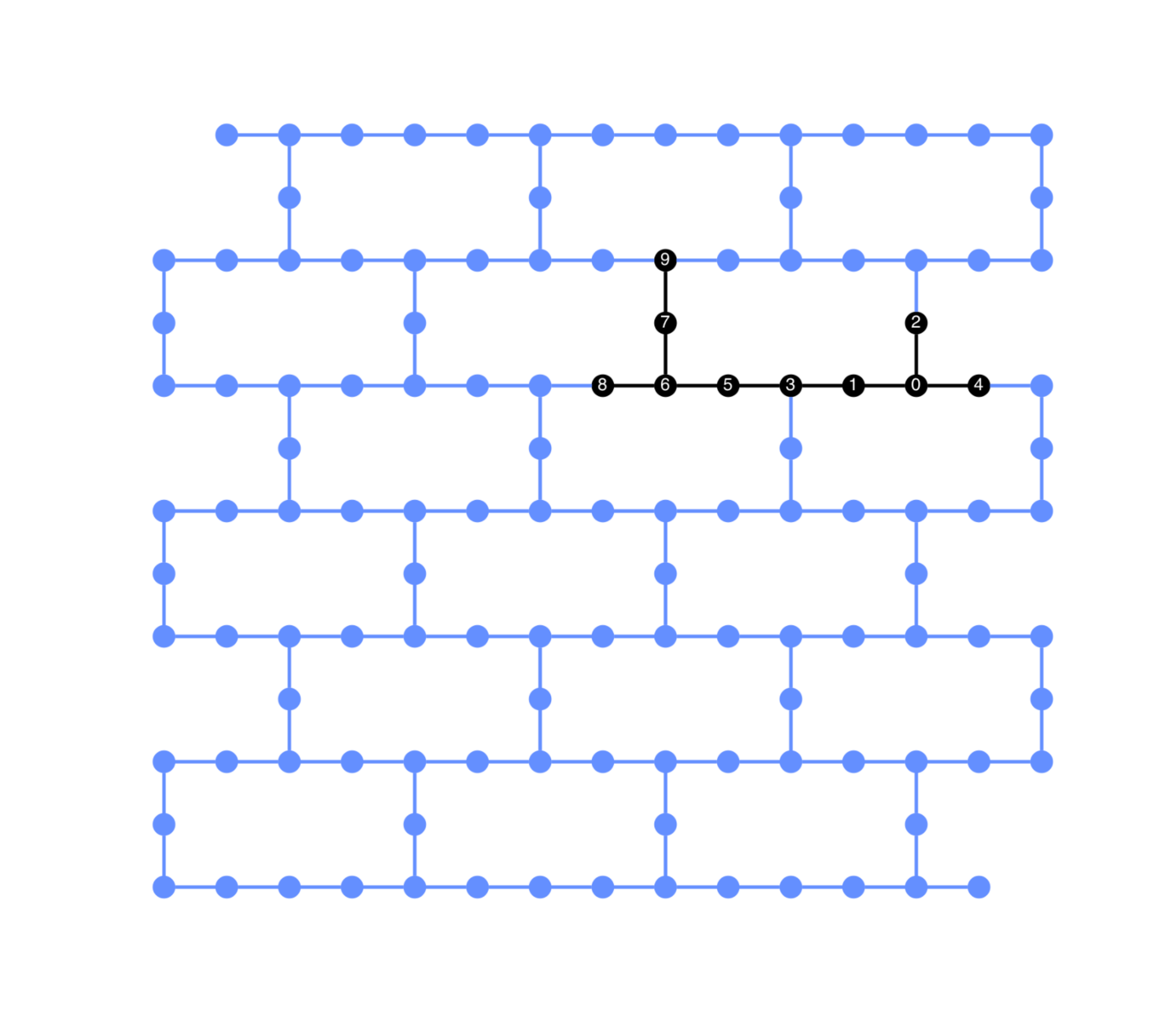}
    \caption{The circuit layout of the IBM quantum computer \textsc{ibm\_strasbourg} for implementing instance W3\_10 is shown for an iteration count of 100, and $k-$stepsize of 4, using one layer ansatz. The used qubits are labeled by numbers and highlighted in black.}\label{fig: qubit-config}
\end{figure}%

The number of ansatz layers is a critical factor in the performance of quantum circuits for solving 1dBPP. To clarify the impact of circuit depth, we present additional results in the Appendix (see Fig.~\ref{fig: layer}). The step size $k$ is set to $2.5$ with 100 iterations. As the number of ansatz layers increases, both FPS and IPS solutions decrease.
However, QAOA consistently fails to capture all exact partial solutions. In contrast, DC-QAOA and the CD-mixer ansatz efficiently maintain the number of FPS while reducing unfavorable ones, demonstrating the capability of CD terms to refine the solution space.

Although CD-inspired ansatz schemes show a slight improvement over standard QAOA, they still fall short of encompassing all exact partial solutions within a given step size. Moreover, increasing the number of layers can lead to diminished performance, likely due to optimization challenges in the cost landscape. As circuit depth grows, the optimization process may struggle to align with all feasible weight sum configurations, emphasizing the need for more efficient parameter tuning and improved training strategies.\\

\section{Experimental Details on IBM Quantum Computer}
\label{sec: IBM detail}

In Sec.~\ref{sec: exp}, we implement the quantum circuits on IBM's 127-qubit \textsc{ibm\_strasbourg} quantum computer using W3\_10  as an example in Fig.~\ref{fig: qubit-config}. This platform supports five basis gates:
\begin{itemize}
    \item \emph{ECR} (Echoed Cross Resonance): A maximally entangling two-qubit gate with a median error rate of $8.715 \times 10^{-3}$. ECR is equivalent to a CNOT gate, up to single-qubit pre-rotations. Its matrix form is given as:
    \begin{equation}
        ECR = \frac{1}{\sqrt{2}}
    \begin{bmatrix}
    0 & 1 & 0  & i \\
    1 & 0 & -i & 0 \\
    0 & i & 0 & 1 \\
    -i & 0 & 1 & 0
    \end{bmatrix}.
    \end{equation}

    \item \emph{RZ}: A single-qubit rotation around the Z-axis, parameterized by $\theta$, with matrix form:
    \begin{equation}
        RZ(\theta) =
    \begin{bmatrix}
    e^{-i\theta/2} & 0 \\
    0 & e^{i\theta/2}
    \end{bmatrix}.
    \end{equation}

    \item \emph{SX}: The $\sqrt{X}$ gate with a median error rate of $2.573 \times 10^{-4}$. Its matrix form is:
    \begin{equation}
        \emph{SX} = \frac{1}{2}
    \begin{bmatrix}
    1 + i & 1 - i \\
    1 - i & 1 + i
    \end{bmatrix}.
    \end{equation}

    \item \emph{X}: The Pauli-X gate, which flips the state of a single qubit. Its matrix form is:
    \begin{equation}
        X =
    \begin{bmatrix}
    0 & 1 \\
    1 & 0
    \end{bmatrix}.
    \end{equation}
\end{itemize}

At the time when the experiments were conducted, the system had a median relaxation time, $T_1$, of $\SI{296.61}{\micro s}$ and a median dephasing time, $T_2$, of $\SI{192.85}{\micro s}$. The median readout error rate is $1.730 \times 10^{-2}$, and when using ten qubits, the two-qubit gate error rate is $1.18 \times 10^{-2}$. These specifications underline the reliability of this platform for implementing our quantum circuits.

\end{appendices}

\bibliography{sample}
\clearpage

\end{document}